\documentstyle[preprint,aps]{revtex} 
\tightenlines
\begin{document} 
\draft

\title{\begin{flushright}
          {\small IFT-P. 023/98 \,\, gr-qc/9804066}
       \end{flushright}
      Interaction of Hawking radiation and a static electric charge} 


\author{Lu\'\i s C.B. Crispino$^{1,2}$, Atsushi Higuchi$^3$, 
       and George E.A. Matsas$^1$}
\address{$^1$Instituto de F\'\i sica Te\'orica, 
         Universidade Estadual Paulista,\\
         Rua Pamplona 145, 01405-900, S\~ao Paulo, S\~ao Paulo, Brazil\\ 
         $^2$Departamento de F\'\i sica, Universidade Federal do Par\'a,\\ 
         Campus Universit\'ario do Guam\'a, 66075-900, Bel\'em, Par\'a,
         Brazil\\
         $^3$Department of Mathematics, University of York,\\ 
         Heslington, York YO1 5DD, United Kingdom}
\date{\today}
\maketitle 


\begin{abstract}
We investigate whether the equality found for the response of
static scalar sources interacting 
(i) with {\em Hawking radiation in Schwarzschild
spacetime} and 
(ii) with {\em the Fulling-Davies-Unruh thermal bath in the Rindler
wedge} is maintained in the case of electric charges. 
We find a finite result in the Schwarzschild case, which is computed exactly,
in contrast with 
the divergent result associated with the infrared catastrophe 
in the Rindler case, i.e.,
in the case of 
uniformly accelerated charges in Minkowski spacetime.  Thus, the
equality found for scalar sources does not hold for electric charges. 
\end{abstract} 

\narrowtext 




It is well known that the equivalence 
principle played a crucial 
role in the development of General Relativity.  It 
continues to be tested  
with great success \cite{TEP}. Recently many authors have 
asked the question whether or not a quantum version of the 
equivalence principle could
be formulated (see, e.g., Ref.\ \cite{CS}). The main problem in 
accomplishing this task stems from the fact that states in quantum mechanics
are defined globally while the equivalence principle involves only local
quantities.
Hence, only those phenomena which are
characterized by frequencies much higher than  
the spacetime curvature 
are expected to show some equivalence 
for flat and curved spacetimes. 
(For example, Hawking radiation can be derived
by requiring that physics near the black hole
horizon be the same as that in Minkowski spacetime
in the infinite frequency limit.)
Thus, there is no {\it a priori} reason to expect any equivalence for 
low-frequency quantum phenomena in flat and curved spacetimes. 
Very recently, however, an interesting equality 
was found for the response of scalar sources \cite{HMS2}. 
Namely, the response rate of a static point source $q$ in 
Schwarzschild spacetime (with the Unruh vacuum)
is equal to the response rate of the same static source in Rindler  
spacetime (with the Minkowski vacuum), which is
\begin{equation}
R^S=\frac{q^2 a}{4\pi^2} \; , 
\label{main}
\end{equation}
provided that both sources have the same proper acceleration $a$.
(We recall that
a static source in Rindler spacetime is nothing but a uniformly
accelerated source in Minkowski spacetime.)  
The response of structureless static sources is entirely
due to the emission and absorption
of ``{\em zero}-frequency particles.'' Thus, this equality clearly involves 
low frequencies.

The choice of the quantum vacuum
is crucial for this equality \cite{HMS3}.  
For instance, it would not be valid if we replaced the Unruh  
vacuum by the Hartle-Hawking vacuum.
However, since the Unruh vacuum is more physical in the sense that it
corresponds to the quantum state for
a black hole formed by gravitational collapse, 
this equality might be pointing
to some underlying equivalence principle. 
Thus,  it is interesting to see whether or not this equality holds
for other fields.

In this Letter we investigate the response of
a static electric charge in Schwarzschild spacetime interacting with 
photons of Hawking radiation (with the Unruh vacuum).
We find that the response rate is finite.
This immediately implies that there is no equality analogous to that
found in the scalar case because the corresponding rate in Rindler 
spacetime is infrared divergent.  Hence, there seems to be no
deep physical principle behind the result for the scalar case.
We present 
the exact response rate for the electric charge instead of merely showing
that it is infrared finite since 
it is a physically meaningful quantity which
one could measure in principle. 
We first proceed to the quantization of the Maxwell field in the exterior 
region of Schwarzschild spacetime in some detail. 
Then we present the response rate and discuss it.
We use natural units $c = \hbar = G = k_B = 1$ and signature $(+---)$ 
throughout this Letter.

The line element for the exterior 
region of Schwarzschild spacetime ($r>2M$)
is given by
$$
ds^2 = f dt^2 - f^{-1} dr^2 - r^2 d\theta ^2 - r^2 \sin^2{\theta} d\phi^2, 
$$
where $f(r) = 1 - 2 M/r$.
We will be interested in a static charge with the current density
of the form $j^\mu = (j^t(r,\theta,\phi), 0, 0 ,0)$. 
However, direct use of this current density would lead to
indefinite results \cite{HMS}. For this reason, we start with an oscillating 
dipole satisfying current conservation, $\nabla_{\mu} j^{\mu} = 0$ :
\begin{equation}
j^\mu =
	(j^t , j^r, 0 ,0) ,
\label{jmu}
\end{equation}
$$
j^t =
	\frac{\sqrt{2}\, q \, \cos {E t}}{r^2 \, {\sin {\theta_0}}}
	 \; [ \delta (r-r_0) - \delta (r-L)] \; 
	 \delta (\theta - \theta_0) \; \delta (\phi - \phi_0) ,
$$
and
$$
j^r =
	\frac{\sqrt{2}\, q\, E \, \sin {E t}}{r^2 \, {\sin {\theta_0}}}
        \; \Theta (r-r_0) \; \Theta (-r+L) \; 
        \delta (\theta - \theta_0) \; \delta (\phi - \phi_0) . 
$$
Here, $\Theta(x) = 1$ if $x > 0$ and $\Theta(x) = 0$ if $x < 0$.
At the end, we take the limit, $L \to \infty$ and $E \to 0$,
to obtain a structureless static point charge
at $(r,\theta,\phi) =(r_0, \theta_0, \phi_0)$.
The normalization of the current has been chosen so that the time average
of the squared charge, 
$(\int d\Sigma_\mu j^\mu)^2$, equals $q^2$.
(A similar normalization was chosen for the scalar case.)

In order to quantize the Maxwell field, we use the standard Lagrangian
density with a covariant gauge-fixing term,
$$
{\cal L} = - {\sqrt -g} \left[ \frac{1}{4} F_{\mu \nu} F^{\mu \nu}
           + \frac{1}{2 \alpha} \left( \nabla^{\mu} A_{\mu} \right) ^2
           \right] .
$$
The corresponding equations of motion in the Feynman gauge $(\alpha=1)$ are
\begin{equation}  
   \nabla_{\nu} \nabla^{\nu} A_{\mu} 
   = 0 .
\label{EqMoFG}
\end{equation}
We write positive-frequency solutions to Eq.\ (\ref{EqMoFG})
with respect to the Killing field $\partial_t$ in the form
$$
	A_{\mu}^{(n,\lambda,\omega,l,m)} = 
	   \zeta_{\mu}^{(n,\lambda,\omega,l,m)} (r, \theta, \phi)
	e^{-i\omega t} ,\ \ \omega > 0,
$$
where we let $n$ = $\rightarrow$ 
for the modes incoming from the past event horizon and
$n$ = $\leftarrow$ for those incoming from the past null infinity. 
The $l$ and $m$ are the angular momentum quantum numbers. 
The label $\lambda$ is for the four polarizations. 
The pure gauge modes with $\lambda = {\rm G}$ are 
the modes which can be written as
$A_{\mu}^{(n,{\rm G},\omega ,l,m)} = \nabla_{\mu} \Phi$ for some  
scalar field $\Phi (x)$ and satisfy Lorenz condition, 
$\nabla^{\mu} A_{\mu}^{(n,{\rm G},\omega,l,m)} =  0$.
The physical modes with $\lambda = {\rm I}$ or II satisfy 
the Lorenz condition, and are not pure gauge.
Finally the nonphysical modes with  $\lambda = {\rm NP}$ do not satisfy
the Lorenz condition.
The Maxwell field operator can be expanded in terms of 
annihilation and creation operators associated with these modes as 
$$
	\hat{A} _{\mu} (x) = \sum_{n,\lambda,l,m} \;
	\int_0^{\infty} d \omega \; \left[ a_{(i)} 
	A^{(i)}_{\mu}(x) 
	+ a^{\dagger}_{(i)} A^{(i) \ast}_{\mu}(x) \right] \; ,
$$
where $(i)$ represents $(n,\lambda,\omega,l,m)$.
We follow the Gupta-Bleuler procedure generalized to curved
spacetime. Thus, we impose the condition 
$\nabla^{\mu} \hat{A}_{\mu}^{(+)}\,| phys \rangle = 0$,
where $\hat{A}_{\mu}^{(+)}$ is the positive-frequency part of the
operator $\hat{A}_{\mu}$,
on the Hilbert space of the physical states.

For the 
sake of brevity, we will just write down the physical modes.
Their derivation will be presented elsewhere.
The modes we call the physical modes
I can be written as
\begin{equation}
	A_\mu ^{(n,{\rm I},\omega,l,m)}=
	    \omega^{3/2}(B_t ^{(n,{\rm I},\omega,l,m)} - 
	    \partial_t \Psi, B_r ^{(n,{\rm I},\omega,l,m)} - \partial_r \Psi,
	    -\partial_{\theta} \Psi,-\partial_{\phi} \Psi) ,
\label{AII} 
\end{equation}
where 
$$
	B^{(n,{\rm I},\omega ,l,m)}_t  = \frac{i}{M \omega } \left[
	(z-1)\frac{d q^n_{\omega l}(z)}{d z} + \frac{(z-1)}{(z+1)} 
	q^n_{\omega l}(z) \right] Y_{lm} (\theta , \phi ) e^{-i \omega t} ,
$$
$$
	B_r^{(n,{\rm I},\omega,l,m)} 
= \frac{(z+1)^2}{(z-1)} q^n_{\omega l}(z) 
	Y_{l m} (\theta , \phi) e^{-i \omega t} ,
$$
and
$$
\Box_s \Psi = - \frac{2 f}{r} B_r^{(n,{\rm I},\omega,l,m)} .
$$
Here, $z \equiv r/M -1$, the $Y_{l m} (\theta , \phi)$ are 
the usual spherical harmonics
and $l\geq 1$. The operator $\Box_s$ is the Laplace-Beltrami operator for
the scalar field.
The $q_{\omega l}^n(z)$ are solutions of the differential equation
\begin{equation}
	\frac{d}{dz} \left[ (1-z^2) \frac{d q_{\omega l}^n}{dz}  \right] + 
	\left[ l(l+1) - \frac{2}{z+1} - M^2 \omega^2 \frac{(z+1)^3}{(z-1)}
	\right] q_{\omega l}^n = 0 .
\label{qwl}
\end{equation}
The $q_{\omega l}^\rightarrow (z)$ satisfy the boundary codition
$q_{\omega l}^\rightarrow (z) \sim e^{iM \omega z}/z$ 
as $z \to \infty$.  On the other hand, 
the $q_{\omega l}^\leftarrow (z)$ satisfy
$q_{\omega l}^\leftarrow (z) \sim (z-1)^{-2iM\omega}$ as $z \to 1$. 
The $l=0$ solutions here 
can be shown to 
be pure gauge.

The other physical modes, which we call the physical modes II,
can be written in the form
$$
	A_\mu ^{(n,{\rm II},\omega,l,m)}=
(0, 0, A_\theta^{(n,{\rm II},\omega,l,m)} , 
	    A_\phi^{(n,{\rm II},\omega,l,m)}) ,
$$
where $l\geq 1$ and $A_j^{(n,{\rm II},\omega,l,m)}\propto 
        (z+1) q^n_{\omega l} (z) 
        Y^{(lm)}_j (\theta , \phi)  e^{-i\omega t} \; $,
$j = \theta , \phi$.
The $Y^{(lm)}_j (\theta , \phi)$ are the divergence-free
vector spherical harmonics
(see, e.g., Ref.\ \cite{AHCQG}).

The normalization factors for the functions $q_{\omega l}^n(z)$
are determined from the canonical
commutation relations of the fields by  requiring suitable commutation 
relations for the annihilation and creation operators. It is 
convenient in this context to define the  generalized 
Klein-Gordon inner product,
\begin{equation}
	\left( A^{(i)}, A^{(j)} \right) \equiv \int_{\Sigma} d\Sigma_{\mu} 
	W^{\mu} [A^{(i)}, A^{(j)}],
\label{AA}
\end{equation}
between any two modes $A^{(i)}_\mu$ and $A^{(j)}_\mu$, where the integration is 
performed on some Cauchy surface $\Sigma$ .
Here,
\begin{equation}
	W^{\mu}	[A^{(i)}, A^{(j)}] \equiv \frac{i}{\sqrt {-g}}
	(A_{\nu}^{(i)\ast} \pi^{(j)\mu \nu} - A_{\nu}^{(j)} 
	\pi^{(i)\mu \nu \ast} ) ,
\label{WmuGI}
\end{equation}
with $\pi^{(i)\mu \nu} \equiv \left.
	\partial {\cal L}/ \partial [ \nabla_{\mu} A_{\nu} ]
	\right| _ {A_\mu = A^{(i)}_{ \mu}}
	= - {\sqrt {-g}} \left[ F^{\mu \nu} + g^{\mu \nu}
	\left( \nabla^{\beta} {A_{\beta}} \right)
	\right]_ {A_{\mu} = A^{(i)}_{\mu}}$ .
It can be shown
that the field equations ensure conservation 
of the current (\ref{WmuGI}), and that
the inner product (\ref{AA}) is independent
of the choice of the Cauchy surface $\Sigma$ as a consequence (see,
e.g., Ref.\ \cite{Fri}). 
Moreover, the inner
product (\ref{AA}) is gauge invariant for physical (and pure gauge) 
modes.  As has been pointed out elsewhere \cite{AtNuPh}, 
the canonical commutation relations among fields and their conjugate 
momenta lead to those of the annihilation and creation operators given
schematically as
$\left[ a_{(i)} , a_{(j)} \right] = \left[ a^{\dagger}_{(i)} , 
	a^{\dagger}_{(j)} \right] = 0$	
and
$\left[ a_{(i)} , a^{\dagger}_{(j)} \right] = \left( M^{-1}
	\right)_{(i)(j)}$,
where $M^{(i)(j)} \equiv \left(A^{(i)}, A^{(j)}\right)$.
The pure gauge and nonphysical modes can be chosen to be orthogonal to the
physical modes with respect to the inner product (\ref{AA}). 
Thus, by requiring the usual 
commutation relations for annihilation and creation
operators, 
$$
	\left[ a_{(n,\lambda,\omega,l,m)} , 
	a^{\dagger} _{(n',\lambda' ,\omega' ,l',m')} \right] = 
	\delta_{nn'}\delta_{\lambda \lambda'}  
	\delta_{l l'} \delta_{m m'} \delta (\omega - \omega') 
$$
with $\lambda$, $\lambda'$ corresponding to the physical modes I and II, 
we are led to the following normalization condition:
\begin{equation}
	\left( A^{(n,\lambda,\omega,l,m)}, A^{(n',\lambda' ,\omega' ,l',m')}
	\right) =\delta^{nn'}\delta^{\lambda \lambda'} 
	\delta^{l l'} \delta^{m m'} \delta (\omega - \omega') \; . 
\label{AAdelta}
\end{equation}

The classical
electric charge interacts with the Maxwell field via the interaction
Lagrangian density
$$
	{\cal L}_{int} = {\sqrt {-g}} \; j^{\mu} A_{\mu} .
$$
Recall that the thermal bath of photons come entirely from the past
event horizon in the Unruh vacuum.  Therefore, we need to consider only
the modes with $n$ = $\rightarrow$.  
Note also that only the physical modes I are 
excited by the current (\ref{jmu}), 
because $A_t = A_r = 0$ for the physical modes II,
once the nonphysical
modes are appropriately chosen.

The particle emission probability with fixed angular momentum for a 
static charge at $(r_0,\theta_0,\phi_0)$ immersed in the Hawking
radiation with temperature $\beta^{-1} = 1/(8\pi M)$ is 
\begin{equation}
	{\cal P}^{em}_{l m} = \lim_{L \rightarrow + \infty} \; 
	\lim_{E \rightarrow 0} \; \int^{+ \infty}_0 d\omega \;
	|{\cal A}^{em}_{(\rightarrow,{\rm I},\omega , l,m)}|^2 \left[ 1 + 
	\frac{1}{e^{\omega \beta} - 1} \right] \, ,
\label{Pem}
\end{equation}
where
$$
	{\cal A}^{em}_{(\rightarrow,{\rm I},\omega , l,m)} 
= \; \langle \rightarrow,{\rm I},\omega , l,m |
	\; i \int d^4 x {\sqrt {-g}} \; j^{\mu} (x) \hat{A} _{\mu} (x) \;
	| 0 \rangle  
$$
is the (Boulware) vacuum emission amplitude of a
photon, in the lowest order of perturbation theory. 
Note that in the static charge
limit, $L \to \infty$ and $E\to 0$, the current will 
interact only with zero-energy modes.
Hence, we need only
the functions $q_{0l}^\rightarrow (z)$, which are the $\omega\to 0$ limit of 
the solutions $q_{\omega l}^\rightarrow (z)$ to Eq.\ (\ref{qwl}).
The normalization factor determined by (\ref{AAdelta}) can be calculated 
by the procedure used in the scalar case \cite{HMS2}.  Thus, we find 
\begin{equation}
q_{0l}^\rightarrow (z) = \frac{2M}{\sqrt{\pi l(l+1)}}
\left[ Q_l(z) - \frac{z-1}{l(l+1)}\frac{dQ_l(z)}{dz}\right],
\label{q0l}
\end{equation}
where the $Q_l(z)$ are the Legendre functions of the second kind. 
(Gauge invariance of the inner product (\ref{AA}) allows us to use
$B^{(n,{\rm I}, \omega,l,m)}$ in place of
$A^{(n,{\rm I}, \omega,l,m)}$ in determining the normalization factor.)
By substituting (\ref{q0l}) in (\ref{Pem}) and using the
differential equation satisfied by $Q_l(z)$, we find 
$$
\frac{{\cal P}^{em}_{lm}}{T} = 
\frac{q^2 (z_0-1)^2}{2\pi M l(l+1) f^{1/2}(r_0)}
\left[\frac{dQ_l(z_0)}{dz_0}\right]^2 |Y_{lm}(\theta_0,\phi_0)|^2,
$$
where $T = 2\pi f^{1/2} (r_0) \delta (0)$ is the total proper time and
where $z_0 = r_0/M-1$.  
One can sum over the angular momentum quantum numbers
$l$ and $m$ by using the formula
$$
\sum_{l=1}^\infty\frac{2l+1}{l(l+1)}\left[
\frac{dQ_l(z)}{dz}\right]^2 = \frac{2Q_1(z)}{(z^2 -1)^2}\, ,
$$
whose derivation will be given elsewhere.  The resulting total
emission rate is
$$
	\frac{{\cal P}^{em}}{T} = \frac{q^2 \, a(r_0)}{4\pi^2} \; 
	Q_1 \left( \frac{r_0}{M} - 1 \right) \, ,
$$
where $a(r_0) = M r^{-2}_0 f^{-1/2} (r_0)$   
is the proper acceleration of the charge.
Similarly, the particle absorption rate with fixed
angular momentum is 
$$
	{\cal P}^{abs}_{l m} = 
	\lim_{L \rightarrow + \infty} \; 
	\lim_{E \rightarrow 0} \; \int^{+ \infty}_0 d\omega \;
	|{\cal A}^{abs}_{(\rightarrow,{\rm I},\omega , l,m)}|^2 
	\frac{1}{e^{\omega \beta} - 1}  \, ,
$$
where 
$|{\cal A}^{abs}_{(\rightarrow,{\rm I} ,\omega , l,m)}| = 
|{\cal A}^{em}_{(\rightarrow,{\rm I} ,\omega , l,m)}|$ by unitarity. 
As a result, the total response rate of the charge is
\begin{eqnarray}
R^V &\equiv& \frac{{\cal P}^{em}}{T} + \frac{{\cal P}^{abs}}{T}
\nonumber \\
&=& \frac{q^2 \, a(r_0)}{2\pi^2} \; Q_1 \left( \frac{r_0}{M} - 1 \right) \, .
\label{R}
\end{eqnarray}
By recalling that
$$
	Q_1(z) = 
	 \frac{z}{2} \, \ln {\frac{z+1}{z-1}}  - 1 \; ,
$$
it is easy to see that the response rate 
(\ref{R}) diverges as the charge approaches
the horizon and vanishes like $r_0^{-4}$ as $r_0 \to \infty$.
Near the horizon we find 
$R^V \approx [q^2 a(r_0)/2\pi^2]\ln [4Ma(r_0)]$.  One can show using the
result of Ref.\ \cite{HMS} that,
for a charge with constant acceleration $a$ in Minkowski spacetime, the
infrared divergence in the total response rate 
is given by $(q^2 a/2\pi^2)\ln(\kappa^{-1}a)$
if one introduces
an infrared cut-off $\kappa$ for the
momentum transverse to the direction of 
acceleration. 
Comparison of these two formulas
shows that the finite size of the black hole
acts as an infrared cut-off. 

We have derived the response rate of a static electric charge outside a
Schwarzschild black hole 
interacting with Hawking radiation in the Unruh vacuum.
It differs from the result obtained for a scalar source, Eq.\ (\ref{main}),
by a factor of $2 Q_1(r_0/M-1)$. 
In the scalar case, it was found that 
the response rates of static point sources
in Schwarzschild spacetime (with the Unruh vacuum)
and in Rindler spacetime (with the Minkowski vacuum) 
are equal provided that
these point sources have the same proper acceleration. Obviously this
equality does not hold in the vector case since the response rate of a
static charge in Rindler spacetime with the Minkowski vacuum,
which is nothing but a uniformly accelerated charge in Minkowski spacetime, 
is infrared divergent as we have seen. 

To check our procedure of defining the modes in spherical polar coordinates
and normalizing them through Eq.\ (\ref{AA}),
we have used it to calculate the response rate of the dipole
(\ref{jmu}) immersed in a background thermal bath in Minkowski 
spacetime. We numerically verified that it reproduces the standard result 
\cite{IZ}.
This would also be an interesting test for the procedure used in the
quantization of the Maxwell field in Schwarzschild spacetime 
with the gauge $A_0 =0$ recently discussed in Ref.\ \cite{CL}.

Finally, we note that our results are in agreement
with the widely accepted conclusion
in classical electrodynamics
that static charges in gravitational fields do not radiate
\cite{R,B,S}. This is so because the zero-frequency modes
which couple to the static charge considered here do 
not carry energy and,
consequently, cannot be identified with classical radiation. 

\acknowledgments
LC and GM would like to acknowledge partial financial 
support from CAPES through the PICDT program
and Conselho Nacional de Desenvolvimento Cient\'\i fico e
Tecnol\'ogico, respectively.

\end{document}